# Filter banks and the "Intensity Analysis" of EMG

***Abstract***
Vinzenz von Tscharner (2000) has presented an interesting mathematical method for analyzing EMG-data called "intensity analysis". Basically the method is a sort of bandpassing of the signal. The central idea of the method is to describe the "power" (or "intensity") of a non-stationary EMG-signal as a function both of time and of frequency. The connection with wavelet theory is that the filter is constructed by rescaling a given mother wavelet using a special array of scales (center frequencies) with non-constant relative bandwidth. Some aspects of the "intensity analysis" method may seem a bit "ad hoc" and we have therefore undertaken a closer mathematical investigation, showing the connection with the conventional wavelet analysis and giving a somewhat simplified formulation of the method using Morlet wavelets. It is pointed out that the "intensity analysis" method is related to the concept of an equalizer. In order to illustrate the method we apply it to non-stationary EMG-signals of a dynamic leg-extension force-velcity tests.

### *1. The Tscharner filter bank*

One the central problems of signal analysis is to describe the frequency content of a non-stationary signal. A conventional method is that of the Short Time Fourier Transformation. During the last decade wavelet methods have also gained increasing popularity in the biomedical community (Akay 1998). For wavelets the frequency corresponds to the concept of a scale. Von Tscharner (2000) has presented a variation on the wavelet theme specially suited for investigating how the "power" of the EMG-signal in a given "frequency bands" varies with time. The "intensity analysis" method proposed by Tscharner is based on a set of wavelets (known as the *Cauchy/Paul/Poisson wavelets*) defined in frequency space by

$$(1) \quad \hat{\psi}(f_c, scale, f) = \left(\frac{f}{f_c}\right)^{\eta} e^{\left(1-\frac{f}{f_c}\right)\eta} \cdot \Theta(f)$$

with

$$\eta = scale \cdot f_c$$

The function (1) is restricted to non-negative frequencies, $f \geq 0$, by the *Heaviside function* $\Theta(f)$ which is equal to 1 exactly when $f \geq 0$ and 0 otherwise. The function (1) is thus a so called *progressive* or *prograde* wavelet. In Tscharner's approach the array of the *center frequencies* $f_c$ is given by

$$(2) \quad f_c = \frac{1}{scale}(q+j)^r$$

Corresp. author F Borg (borgbros@netti.fi) Biosignals Project, Jyväskylä University, Chydenius Institute

for $j = 0, 1, \ldots, J$. Fixed parameter values chosen by Tscharner are

   $q = 1.45$
   $r = 1.959$
   $scale = 0.3$

(The "scale" parameter here is not to be confused with the the general scale concept in wavelet theory.) With these values the summation of the eleven ($J = 10$) first wavelets in frequency space gives a function that is almost constant in the interval of 20 to 200 Hz (this is more precisely so if one chooses $r = 2$, a reason for this choice is discussed below). The wavelets can thus be used as a filter bank for decomposing signals into frequency bands in this interval (for larger $J$ we get a bigger interval). That the sum of (1) is nearly constant ($C$) in this interval,

(3) $$\sum_j \hat{\psi}_j(f) \approx C$$

seems to imply that the corresponding sum in time-space approximates the Dirac delta (the inverse Fourier transform of the constant unit function). However, due to the restriction to the non-negative frequencies ($f \geq 0$) we have instead[1]

(4) $$\sum_j \psi_j(t) \approx C \left( \frac{\delta(t)}{2} + \frac{i}{2\pi t} \right)$$

(here we have written $\hat{\psi}_j(f)$ for $\hat{\psi}(f_{cj}, scale, f)$ etc; a "hat" over a function denotes here its Fourier-transform, and a "bar" its complex conjugate). In this sense a signal can be approximately deomposed into a sum, according to ($\Re(z)$ stands for real part of a complex number $z$)

(5) $$x_j(t) = \frac{1}{C} \int \bar{\psi}_j(u-t) x(u) du$$
$$x(t) \approx 2 \sum \Re(x_j(t))$$

Thus, the real part of $x_j(t)$ gives a description how the "component" of $x(t)$ centered on the frequency $fc_j$ behaves with time. In the "intensity analysis" proposed by von Tscharner one is rather interested in tracking the "intensity" $p_j(t)$ for the components, which we define here as (Tscharner's definition given in equ (11) below),

(6) $$p_j(t) = |x_j(t)|^2$$

If we sum (6) over $j$ and integrate over time we get (via Parseval's relation)

---

[1] The rhs is (in distributional sense) for the case when the frequency interval becomes infinite. It is the inverse Fourier transformation of the Heaviside function. For mathematical details see e.g. Saichev and Woyczyñski (1997).

(7)     $\sum_j \int p_j(t)\,dt = \int \sum_j |\hat{\psi}_j(f)|^2 \, |\hat{x}(f)|^2 \, df$

and this is proportional to the "energy"    $\int |\hat{x}(f)|^2 \, df$    of the signal if we may assume that

(8)     $\sum_j |\hat{\psi}_j(f)|^2$

too is almost constant in the frequency interval. Thus, for "intensity analysis" it is rather the constancy of sqaured sum (8) rather than the simple sum (3) that is primary. With these points in mind (6) may be interpreted as a measure of the "power" of the signal around the time $t$ and the frequency $fc_j$.

A sinus-wave of frequency $f_0$ whose representation in frequency space is (δ is the Dirac "function")

(9)     $\dfrac{1}{2i}\left(\delta(f - f_0) - \delta(f + f_0)\right)$

will be transformed by the wavelet (1) to (in time space)

(10)    $\dfrac{1}{2i} \cdot \left(\dfrac{f_0}{f_c}\right)^\eta \cdot e^{\left(1 - \frac{f_0}{f_c}\right)\eta + i 2\pi f_0 t}$

The output is thus a complex function whose magintude is constant. Von Tscharner (2000, p. 438) describes a somewhat intricate method for calculating the "intensity" for a transformed signal $v$. If $v_j$ is the real transform of the signal by the wavelet $j$ then he calculates the "intensity" $p_j$ as

(11)    $p_j(t) = v_j(t)^2 + \left(\dfrac{1}{2\pi fc_j} \dfrac{dv_j(t)}{dt}\right)^2$

The last term on the RHS in (11) is included according to von Tscharner in order to get rid of the oscillatory terms when the transform is e.g. applied to a sinus-wave. However, this extra term is not needed if we use both the real and imaginary parts of the wavelet (9) and we take the magnitude of the complex transformed function as a measure of the intensity. The method (11) will give the same result in case of the infinite (stationary) sinus-wave but in other cases only approximately so.

In time space the wavelet (1) becomes a complex function (Γ denotes the gamma function)

Corresp. author F Borg (borgbros@netti.fi) Biosignals Project, Jyväskylä University, Chydenius Institute

(12)
$$\psi(f_c, \eta, t) = \Gamma(\eta+1) e^{\eta} \frac{f_c}{(\eta - i2\pi f_c t)^{\eta+1}}$$
with
$$\eta = scale \cdot f_c$$

The frequency resolution for (1) can be calculated to be

(13)    $\Delta f = fc \cdot \sqrt{\frac{1}{2\eta}\left(1 + \frac{1}{2\eta}\right)}$

and the mean frequency

(14)    $\langle f \rangle = fc \cdot \left(1 + \frac{1}{2\eta}\right)$

while (1) attains the maximum value 1 at $f = f_c$. The corresponding time resolution of (9) is found to be (only defined for $\eta > 0.5$)

(15)    $\Delta t = \frac{1}{2\pi f_c} \frac{\eta}{\sqrt{2\eta - 1}}$

Thus, the time-frequency "uncertainty" relation becomes

(16)    $\Delta f \cdot \Delta t = \frac{1}{4\pi} \cdot \sqrt{\frac{2\eta+1}{2\eta-1}}$

which is very close to the optimal result $1/4\pi$ when $\eta \gg 1$.

## *2. A Morlet filter bank*

One can obtain a similar filter bank using Morlet "wavelets"

$$(17) \quad \hat{\psi}(f_c, \alpha, f) = e^{-\frac{2\pi^2}{\alpha f_c}(f - f_c)^2}$$

Compared to the standard form of the Morlet function the parameter α is here multiplied with $f_c$ (see further § 3). The summation of these wavelets is shown in fig. 1 using α = 150 and the center frequencies (2) and $J = 10$.

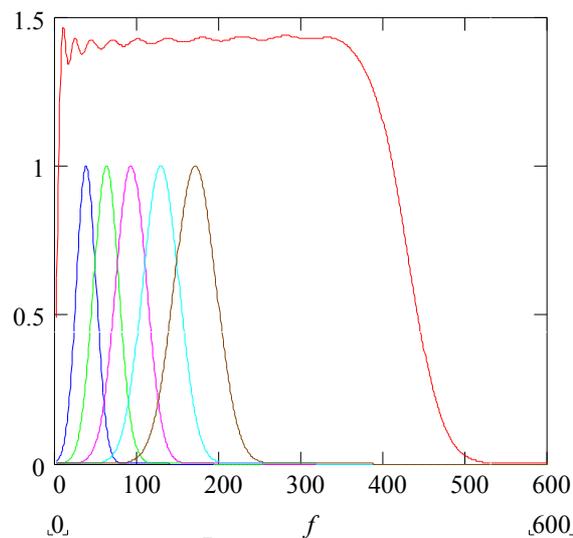

Fig. 1. The graphs show wavelets (17) for j = 2,...,6 and their sum from j =0 to j = 10.

The time and frequency resolutions for the Morlet (17) are given by

$$(18) \quad \Delta t = \frac{1}{\sqrt{2\,\alpha f_c}}$$

$$(19) \quad \Delta f = \frac{\sqrt{2\,\alpha f_c}}{4\pi}$$

and are thus optimal in term of the time-frequency relation. These relations agree approximately with (13) and (15) if we choose α as

$$(20) \quad \alpha = \frac{4\pi^2}{scale} \quad \text{when} \quad \eta = scale \cdot f_c \,.$$

Corresp. author F Borg (borgbros@netti.fi) Biosignals Project, Jyväskylä University, Chydenius Institute

A few of the corresponding Morlet wavelets in time space shown in fig. 6 are given by

$$(21) \quad \psi(t) = \sqrt{\frac{\alpha f_c}{2\pi}} \cdot e^{i 2\pi f_c t - \alpha f_c \frac{t^2}{2}}$$

which is Gaussian unlike (1) (the Paul wavelet (1) approximates though a Gaussian function in frequency space for $f$ close to the center frequency $f_c$). In the present case the time resolution varies from 22 ms to 3 ms and the frequency resolution from 3.6 Hz to 36.4 Hz ($j$ = 0, ..., 10). From (19) it follows that the frequency resolution grows linearly with $j$ if $r$ is close to 2 in (2).

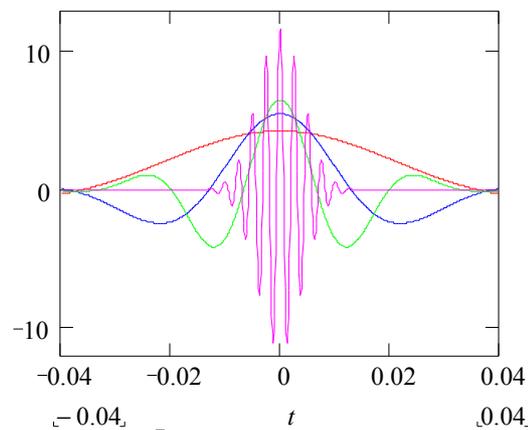

*Fig. 2. Graphs show real part of the Morlet functions (21) for j = 0, 1, 2, 10.*

Strictly speaking the Morlet wavelets are not proper wavelets because they do not satisfy the so called admissibility condition $\hat{\psi}(0) = 0$, still (1) is very small for $f = 0$, so this circumstance may have not much practical implications anyway. The Morlet wavelets and Paul wavelets give within the approximations used here practically identical results, but the Morlet has perhaps some nicer mathematical properties from the computational point of view.

### *3. Comparison with the conventional continuous wavelet transform (CWT)*

The new idea with von Tscharner's method seems to be the use the special set (2) of center frequencies. Also the analysis is not primarily based on an attempt to decompose the signal into a sum of wavelets but to use a filter bank strictly for power analysis. The "standard" Morlet wavelet in frequency space is written (non-normalized)

$$(22) \quad \hat{\psi}(\alpha, f_c, f) = e^{-\frac{2\pi^2}{\alpha}(f - f_c)^2}$$

Note that here the $f_c$-factor is dropped after $\alpha$ in the exponent as compared to (17). Thus, the formulas corresponding to (18-19) become in this case

(23) $\quad \Delta t = \dfrac{1}{\sqrt{2\,\alpha}}$

(24) $\quad \Delta f = \dfrac{\sqrt{2\,\alpha}}{4\,\pi}$

The continuous wavelet transform is usually calculated for scales and frequencies parametrized as ($p > 1$)

(25) $\quad \begin{aligned}\alpha_j &= p^{2j}\alpha_0 \\ f_j &= p^j f_0\end{aligned}$

This will imply, using (24), that the *relative bandwidth BW* defined by

(26) $\quad BW = \dfrac{\Delta f}{f_c}$

remains constant. The previous filter banks differ by having a non-constant relative bandwidth. The Tscharner filter bank has a relative bandwidth which decreases as

(27) $\quad BW \approx \dfrac{1}{\sqrt{scale \cdot f_c}}$

when $scale \cdot f_c \gg 1$. Indeed, in order that the sum

(28) $\quad \sum_j \hat{\psi}(\alpha_j, f_j, f)$

be approximately independent of $f$ in the range of interest, the frequency resolution $\Delta f$ should scale approximately as $f_{j+1} - f_j$; that is, given (2) (with $r = 2$) we should have

(29)   $\Delta f \sim \sqrt{f}$

This is indeed satisfied if we choose the form (17) for the scaling of the Morlet function. We can illustrate the situation employing a much simpler division of the frequency band by using rectangles (see fig. 3).

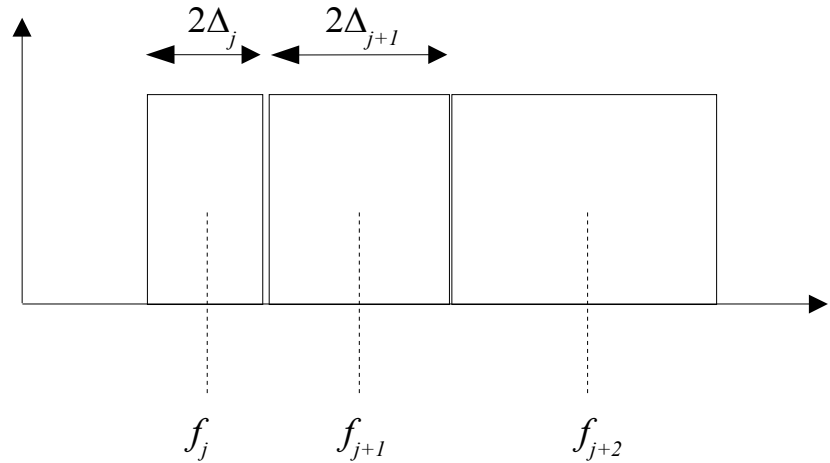

*Fig. 3. A schematic division of the frequency range into frequency bands.*

Thus, suppose we have boxes centered at $f_j$ and with widths $\Delta_j$ related by,

(30)   $f_{j+1} - f_j = \Delta_{j+1} + \Delta_j$

From this it follows that if $f_j$ is a quadratic function of the index $j$,

(31)   $f_j = a(q+j)^2$

then we get for the (half-) box sizes

(32)   $\Delta_j = a(q+j) = \sqrt{a f_j}$

which can be compared with (29). This indicates a general relationship between center-frequencies (31) and widths for filter banks that try to minimize the overlap. Naturally, the box-filter as such is far from ideal when we consider its behaviour in time (the "ringing" sinc-function). The signal $x$ is analyzed by the wavelets (17) as

(33) $\quad c_j(t) = \int \bar{\psi}_j(u-t) x(u) du$

where $\psi_j$ is shorthand for $\psi(\alpha_j, f_j, t)$. In frequency space we get

(34) $\quad \sum |\hat{c}_j(f)|^2 = \sum |\hat{\psi}_j(f)|^2 \cdot |\hat{x}(f)|^2$

Thus, if the sum $\sum |\hat{\psi}_j(f)|^2$ is also approximately constant in the frequency band then (34) will be proportional to the power of the signal at the frequency $f$. The value $|c_j(t)|^2$ may thus be taken as a measure of the power of the signal around frequency $f_j$ at the time $t$. So in this sense the condition of the constancy of the sum (8) of the squared amplitudes of the wavelet in frequency space is required, whereas the condition (3) is only indirectly related to this. Anyway, this condition requires that adjacent wavelets $\psi_j$ and $\psi_{j+1}$ overlap enough so that the wavelets cover the frequency band without gaps. Furthermore, in order that the "intensity" $p_j$ may be interpreted as measure of the "power" in a frequency band centered on $fc_j$, the wavelets $\psi_j$ and $\psi_{j+2}$ should have a minimal overlap and thus separate the different subbands (minimal "leakage" between bands). By using a relative bandwidth of the form (27) these objectives can be approximately fulfilled. Tscharner (2000) provides thus an interesting but a somewhat *ad hoc* recipe.

The use of non-normalized wavelets (1) and (17) brings in a further aspect related to the concept of *equalizer* (EQ) familiar from audio-technology. This refers to the ability to independently amplify the signal in different frequency bands. The conventional continuous wavelet transform (CWT) for a scaling parameter $a > 0$ is defined by

(35) $\quad T(a,t) = \frac{1}{\sqrt{a}} \int \bar{\psi}\left(\frac{u-t}{a}\right) x(u) du$

using a normalization factor $1/\sqrt{a}$. In frequency space (35) becomes

(36) $\quad \hat{T}(a,f) = \sqrt{a}\, \bar{\hat{\psi}}(af)\, \hat{x}(f)$

The scalings (25) correspond to using $a = p^{-j}$ in (35-36); that is, the scaled wavelet in the frequency space is multiplied by a factor

(37) $\quad \sqrt{a} = p^{-\frac{j}{2}}$

whereas in the "intensity analysis" there is no such multiplicative factor, which means that the higher frequency components are enhanced by a factor of $1/\sqrt{a}$ as compared with the conventional wavelet transformation (35). On the other hand the width of the wavelet in (35) scales for the Morlet as $1/a^2$ and for the intensity analysis method as $1/a$ (narrower shape). In all there are thus some characteristic differences in the equalizing by these two methods. Clearly one could develop, in case they turn out to be useful, analyzing methods where the equalizing parameters can be freely adjusted channel by channel.

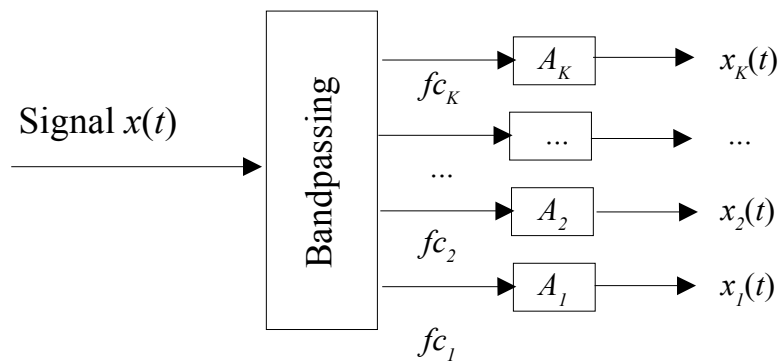

*Fig. 4. A basic "equalizer" setup. Input signal x(t) is "decomposed" into components associated with center frequencies $fc_i$. The components can then be independently amplified, processed, etc.*

## 4. Example: analysis of nonstationary leg-extension EMG-signal

We will make a brief analysis of EMG-data in order to highlight some aspects of the method. The data was obtained in connection with a leg-extension force-velocity (FV) test using a pneumatic resistance device (HUR Research line Leg-extension/curl). In the FV-test one makes a series of MVC extensions with the lower leg, increasing the load between the repetitions. In this way it is possible to map the force-velocity curve. (For a detailed description see Borg and Herrala (2002).) The MVC kick is quite swift with velocities up to 500 degrees/sec. The movement is completed in a time around 0.3 – 0.5 sec depending on the load. The FV-test was performed with a simultaneous recording (1000 S/s) of surface EMG (Noraxon Myosystem 1400 with a 10 Hz high-pass filter) from *Vastus medialis* (VM), *Vastus lateralis* (VL), and *Rectus femoris* (RF). As we can see from the representative graphs of the test we have naturally very short bursts of EMG, which is far from the stationary case expected e.g. in Fourier analysis. Next we will apply the Morlet filter bank version (17) of the "intensity analysis" to the data using the parameters

$\alpha = 150$

$scale = 0.5$

$q = 1.45$

These give the center frequency values (equ (2) with $r = 2$)

| $j$ | $fc_j$ (Hz) | $\Delta f$ |
|---|---|---|
| 0 | 4.2 | 2.8 |
| 1 | 12.0 | 4.8 |
| 2 | 23.8 | 6.7 |
| 3 | 39.6 | 8.7 |
| 4 | 59.4 | 10.6 |
| 5 | 83.2 | 12.6 |
| 6 | 111.0 | 14.5 |
| 7 | 142.8 | 16.5 |
| 8 | 178.6 | 18.4 |
| 9 | 218.4 | 20.4 |

*Table 1. Center frequenices. (When using Matlab note that indexes starts at 1.)*

By changing the parameter values or increasing the number of center frequencies one can go to higher frequency regions, but in the present case there is not much "energy" in the EMG signal for frequencies above 200 Hz.

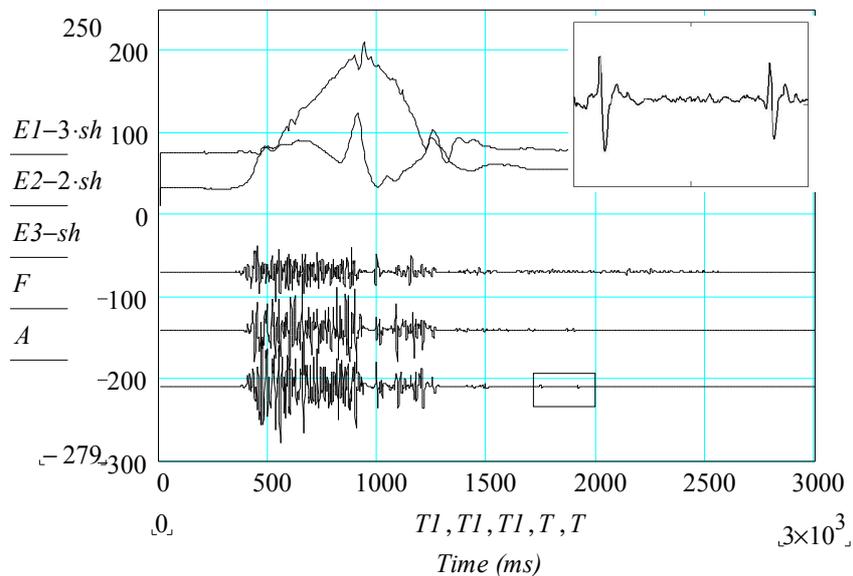

*Fig. 5. Representative data of FV-test ($\bar{3}$d repetition). Upper graph records angle (180 when the participant keeps the leg "straight"), the next graph describes the torque (Nm), and then in order we have raw EMG from RF, VM and VL (bottom). The active extension takes place between the times around 500 – 900 ms. The torque peak around 900 ms is due to the leverarm hitting the stopper. Note the post-motion EMG reflex MUAPs shown enlarged in the insert.*

Coresp. author F Borg (borgbros@netti.fi) Biosignals Project, Jyväskylä University, Chydenius Institute

In fig. 6 are displayed the "intensity" as a function of time of the EMG shown in fig. 5 (VL). Apparently the fourth component centered on the frequency about 60 Hz contains most of the EMG "energy". It is thus possible e.g. to compare how the distribution of the intensity among the components varies with time.

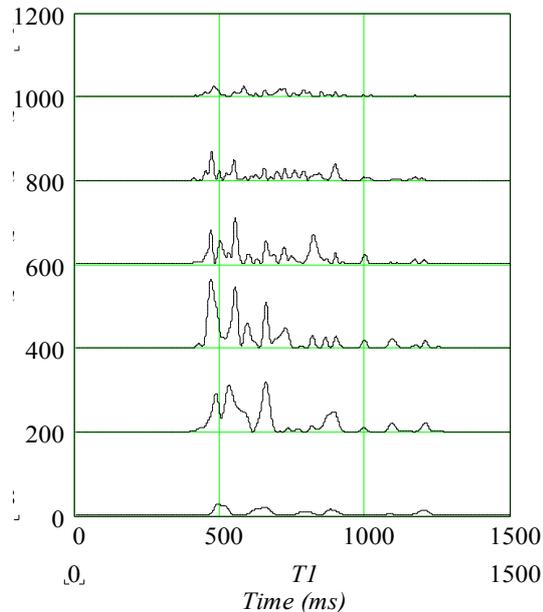

*Fig. 6. The figure shows the "intensity" of the VL EMG-signal of fig. 5 as a function of time, calculated according to equ (6) using the Morlet filter and the center frequencies in Table 1. Shown are the cases j = 2, ..., 7 in the order from bottom up.*

As a simple application we calculated the maximum of the "intensity" components for each repetition (six in all), see fig 7. When we go from bottom curve upwards the load of the MVC-kick increases. As expected the EMG "intensity" increases with the load. The last repetition shows a very large maximum intensity value as compared with the other ones (the participant is obviously "squeezing" all he has got left in the last rep).

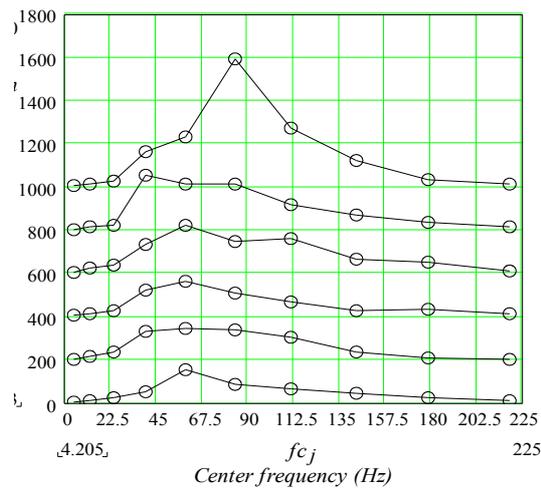

*Fig. 7. The maximum intensities for reps 0-5 in the order from bottom upwards as a function of the center frequencies.*

A major contribution in the last rep comes from the component $j = 5$ corresponding to the center frequency 83 Hz. The maximum intensities may be related to the onset of the "explosive" motion (see fig. 6, $j = 4$ curve).

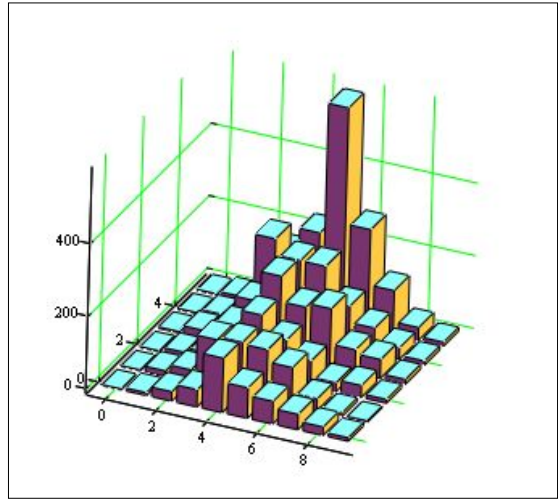

*Fig. 8. The same as fig. 7 but presented as a bar surface graph. From left to right the center frequency indices $j = 0 - 9$. From front backwards repetition 0 (lightest) to repetition 5 (heaviest load).*

Corresp. author F Borg (borgbros@netti.fi) Biosignals Project, Jyväskylä University, Chydenius Institute

## *Appendix. Matlab program for "intensity" calculation.*

% morlet_vt.m: Morlet-vTscharner Filter Bank Intensity Analysis

function [B, Fc]= morlet_vt(X, fs, alpha, J, q, scale);

% Implementation: F Borg (borgbros@netti.fi), Jyväskylä University, Chydenius Institute 2003.

% Based on using fft/ifft for "fast" calculation of the convolution. Tested on Matlab version 6.1.

% INPUT parameters, values:

% X: 1-dim data array [row] of size N

% fs: sampling rate of data (S/s)

% alpha: sqrt(alpha) characterizes the width of the freg. distribution of the Morlet function [typ. value alpha = 150]

% J: number of center frequencies, indexed j = 0, ..., J-1  [typ. value J = 10]

% q: offset parameter that determines the values of the center frequencies  [typ. value q = 1.45]

% scale: scale parameter that determines the values of the center frequencies [typ. value q = 0.5]

% OUTPUT values:

% B: N*J-matrix consisting of J columns (of size N) corresponding to the center frequencies

% Fc: array [row] of the center frequenices

```
    explimit = 20;                                 % if x > explimit set exp(-x) = 0
    N = length(X);
    N2 = 2^nextpow2(X);
    N2_2 = N2/2;
    I = linspace(0, J-1, J);
    Fc = (q + I).*(q + I)/scale;                   % center frequencies accoring to the r = 2 formula
    lambda = 2*pi*pi/alpha;
    FX = fft(X, N2);
    f = linspace(0, N2-1,N2).*fs/N2;               % frequency values
    for j=0:J-1;
       dummy = lambda/Fc(j+1);
       for k=1:N2;
          u = dummy*(f(k) - Fc(j+1))*(f(k) – Fc(j+1));   % obs! Matlab starts indexing arrays at 1, not 0
          if(u > explimit) Q(k) = 0;
          else Q(k) = 1/exp(u);
          end;
       end;
       S = ifft(FX.*Q);                            % convolution calculation using FFT/IFFT
       B(j+1, 1:N) = S(1:N);
    end;
```

The output matrix value $B(j+1, i)$ corresponds to the value of the transform (with the Morlet function (17))

$$(38) \quad c_j(t) = \int \bar{\psi}_j(u-t) x(u) \, du$$

of the data at the time $t$ corresponding to the time index $i$. Note that Matlab (www.mathworks.com) starts indexing vectors and matrices at 1, not 0. In order to calculate the "intensity" one has to take the square magnitude of (38). Thus, in order to plot the intensity vs time for the center frequency $fc_j$ one can use the commands:

```
t = linspace(0, (n-1)/fs, n);              % time points
plot(t, conj(B(j+1, 1:n)).*B(j+1, 1:n));    %plot "intensity" vs time
```

where $fs$ is the sampling rate and $n$ the number of data poins. Calculations were made also by an equivalent implementation in Mathcad (www.mathsoft.com).

## *Acknowledgements*

We are indebted to v Tscharner for personal communications, and discussions during conference meetings in Calgary and Banff 2002 (Canada), on the topics of the "intensity analysis". This paper was written during the project "Biosignals" sponsored by the Finnish technology agency Tekes.

Coresp. author F Borg (borgbros@netti.fi) Biosignals Project, Jyväskylä University, Chydenius Institute